\documentclass[english,nofootinbib,notitlepage,prl,
twocolumn]{revtex4}

\usepackage[T1]{fontenc}
\usepackage[latin9]{inputenc}
\usepackage{color}
\usepackage{graphicx}
\usepackage{amssymb,amsmath}
\usepackage{babel}

\def\be{\begin{equation}}
\def\ee{\end{equation}}

\makeatletter

\@ifundefined{definecolor}
 {\usepackage{color}}{}

\begin{document}

\title{ The Gravitational Wave Background and Higgs False Vacuum Inflation  }

\author{Isabella Masina$^{1,2,3}$}

\affiliation{$^{1}$  Dip.~di Fisica e Scienze della Terra, Universit\`a di Ferrara and INFN Sez.~di Ferrara, Via Saragat 1, I-44100 Ferrara, Italy}
\affiliation{$^{2}$ CP$^3$-Origins \& DIAS, SDU 
Campusvej 55, DK-5230 Odense M, Denmark}
\affiliation{$^{3}$ CERN, Theory Division, CH-1211 Geneva 23, Switzerland}

\begin{abstract}
For a narrow band of values of the top quark and Higgs boson masses, the Standard Model Higgs potential develops 
a shallow local minimum at energies of about $10^{16}$ GeV, where primordial inflation could have started in a cold metastable state. 
For each point of that band, the highness of the Higgs potential at the false minimum is calculable, and there is an associated prediction for the inflationary gravitational wave background, namely for the tensor to scalar ratio $r$. 
We show that the recent measurement of $r$ by the BICEP2 collaboration,  
$r=0.16 _{-0.05}^{+0.06}$ at $1\sigma$,   
combined with the most up-to-date measurements of the top quark and Higgs boson masses,
reveals that the hypothesis that a Standard Model shallow false minimum was the source of inflation in the early Universe is viable.  
 \end{abstract}


 \maketitle



The fact that, for a narrow band of values of the top quark and Higgs boson masses, 
the Standard Model (SM) Higgs potential develops a local minimum \cite{CERN-TH-2683, hep-ph/0104016, strumia2} is very interesting,
as this happens at energy scales of about $10^{16}$ GeV which are suitable for inflation in the early Universe. 

Inflation from a local minimum \cite{Coleman, Guth} is a viable scenario, provided a graceful exit to a radiation-dominated era can be obtained
via some mechanism beyond the SM. 
Developing a model with graceful exit in the framework of a scalar-tensor theory of gravity \cite{hep-ph/0511207,astro-ph/0511396}, 
in ref. \cite{Masina:2011aa} we pointed out that the hypothesis that inflation took place in a SM shallow false vacuum was consistent only with a narrow range of values of the Higgs boson mass, which
subsequently turned out to be compatible with the experimental range indicated by ATLAS and CMS \cite{LHC}.
 
These very suggestive results provide a strong motivation to further 
investigate the scenario of SM false vacuum inflation, by looking for complementary experimental tests.
Inflation can generate tensor (gravity wave) modes as well as scalar (density perturbation) modes. 
It is common to define the tensor contribution through $r$, the ratio of tensor to scalar perturbation spectra at large scales. 
If inflation happened at a very high scale, as is the case for the SM false vacuum scenario, quantum fluctuations during inflation
produced a background of gravitational waves with a relatively large amplitude.

As argued in \cite{Masina:2011un}, the tensor to scalar ratio, combined with the top quark and Higgs boson mass measurements,  
does represent a test of the hypothesis that inflation started from the SM false vacuum. The upper bounds on $r$ provided 
by the WMAP \cite{Komatsu:2010fb} and Planck experiments \cite{Ade:2013uln} were too weak for the sake of such test. 
It was anyway possible to conclude \cite{Masina:2012tz} that, for the SM false vacuum to be a realistic inflationary scenario, an experimental detection of $r$ would have been possible in the case that the top quark mass turned out to be close to its lower allowed value at $2\sigma$.

The recent measurement of $r$ by the BICEP2 collaboration \cite{Ade:2014xna}, $r=0.16\ ^{+0.6}_{-0.5}$ at $1\sigma$, on the one hand represents a hint in favor of the hypothesis that inflation took place in the SM shallow false vacuum and, on the other hand, 
allows to perform a sensible test of the allowed parameter space for such scenario. That is the goal of the present Letter.


Let us consider the Higgs potential in the SM of particle physics.
For very large values of the Higgs field $\chi$, the quadratic term $m^2 \chi^2$ can be neglected and we are left with the quartic term, 
whose dimensionless coupling $\lambda$ depends on the energy scale, which can be identified with the field $\chi$ itself:
\begin{equation}
V(\chi) \simeq \lambda(\chi) \, \chi^4\,\,.
\end{equation}
It is well known that, for some narrow band of the Higgs and top masses, the Higgs potential develops a new local minimum 
\cite{CERN-TH-2683, hep-ph/0104016, strumia2}. 

If the Higgs field is trapped in a cold coherent state in the false minimum $\chi_0$ and dominates the energy density of the Universe, 
the standard Friedmann equation leads to a stage of inflationary expansion 
\be
H^2  \simeq \frac{V(\chi_0)}{3 M^2} \equiv H_I^2 \,\,\,,\,\,\,  \,\,a(t)\propto e^{H_I t}  \, \, ,
\label{eq-M}
\ee
where  $a(t)$ is the scale factor, $H \equiv \dot a /a$ is the Hubble rate and $M=1.22 \times 10^{19}/\sqrt{8 \pi}$ GeV 
is the reduced Planck mass.

A nontrivial model-dependent ingredient is how to achieve a graceful exit from inflation. 
In order to end inflation the Higgs field has to tunnel to the other side of the potential barrier by nucleating bubbles~\cite{Coleman} 
that eventually collide and percolate.
Subsequently the Higgs field could roll down the potential, reheat the Universe and relax in the electroweak vacuum.
A graceful exit can be generically realized only if at the end of inflation there is a very shallow false minimum, 
otherwise the tunneling rate becomes negligibly small, the probability being exponentially sensitive to the barrier \cite{Coleman}.
The shape of the potential is thus very close to an inflection point configuration. 
This leads to a powerful generic prediction for the scale of inflation, and therefore for $r$ \cite{Masina:2011un}. 
So, if the false vacuum is very shallow, the specific  model only affects the prediction for the spectral index of cosmological 
density perturbations $n_S$, see e.g. the models of refs. \cite{Masina:2011aa, Masina:2012yd}.

Using the Renormalization Group Equations (RGE) at Next-to-Next-to-Leading Order (NNLO), {\it i.e.} 3-loops for beta functions \cite{Mihaila:2012fm, Chetyrkin:2012rz} and the 2-loops for matching conditions \cite{Bezrukov:2012sa,Degrassi:2012ry}, we investigated \cite{Masina:2012tz} the values of the top quark and Higgs masses allowing for the presence of a second degenerate minimum or a shallow local minimum at high scale. For the technical details on the RGE procedure at NNLO we refer the interested reader to ref. \cite{Masina:2012tz}, while here we summarize only the issues relevant for the present work. 

The are two main sources of uncertainty in the RGE calculation at NNLO. 
The first is of experimental nature and is associated to the uncertainty in the determination of the value of $\alpha_3(m_Z)$, for which the PDG \cite{PDG} gives the range $\alpha_3(m_Z)= 0.1196 \pm 0.0017$  at  $1\sigma$.  
The second is of theoretical nature and is associated to the matching conditions. As suggested in ref. \cite{Alekhin:2012py}, using the value of the top mass in the $\overline{\rm MS}$ scheme, ${\overline{ m}}_t(m_t)$, one has to consider only the theoretical uncertainty associated to the matching of $\lambda$; 
this kind of top mass is experimentally known with a quite large $1\sigma$ 
error, ${\overline{ m}}_t(m_t)=163.3 \pm 2.7 $ GeV according to \cite{Alekhin:2012py},
and ${\overline{ m}}_t(m_t)=160 ^{+5}_{-4} $ GeV according to the PDG \cite{PDG}. 
Using instead the top pole mass, the theoretical error becomes bigger because one has to include also the uncertainty associated to the matching of the top yukawa coupling (see e.g. \cite{Bezrukov:2012sa,Degrassi:2012ry});  the top pole mass $m_t$ is however known quite accurately, $m_t=173.0\pm1.2$ GeV at $1\sigma$ according to the PDG \cite{PDG}, and $m_t =173.34 \pm 0.76$ GeV according to the 
recent combination of the ATLAS, CDF, CMS, D0 measurements \cite{ATLAS:2014wva}.
In the present analysis, as a crosscheck, we adopt both procedures, finding that the two methods essentially provide similar results (as they should), even though the second method is somewhat more stringent.

\begin{figure}[h!]
 \includegraphics[width=7.5cm]{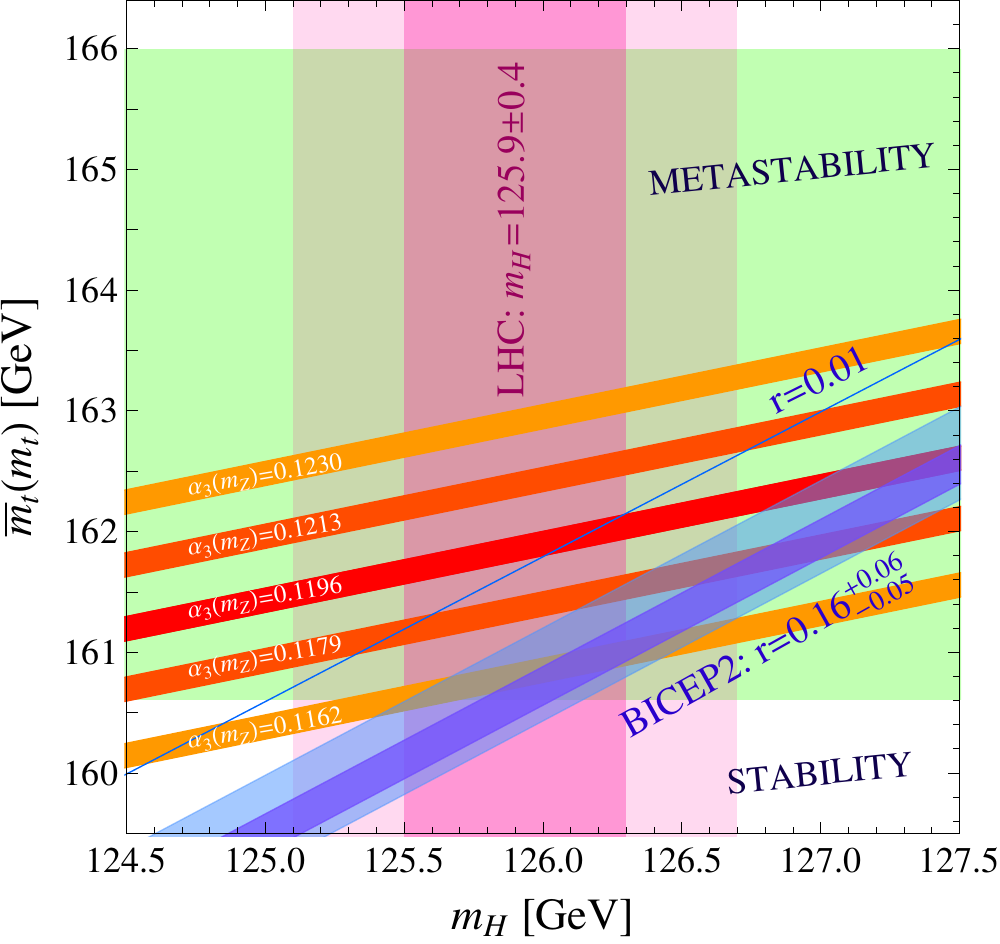}  
\caption{The (red and orange) lines show the points where the SM shallow minimum configuration is realized, their thickness being the theoretical error in the NNLO RGE procedure. Each line corresponds to a value of $\alpha_3(m_z)$, taken to vary in its $2\sigma$ range. 
The (green)  shaded horizontal region corresponds to the $1 \sigma$ range ${\overline{ m}}_t(m_t)=163.3 \pm 2.7 $ GeV \cite{Alekhin:2012py}. 
The (pink) shaded vertical regions are the 1 and 2 $\sigma$ ranges of $m_H$ \cite{PDG}. The (blue) shaded bands are the $1$ and $2$ 
$\sigma$ ranges of BICEP2, $r=0.16\ ^{+0.06}_{-0.05}$ \cite{Ade:2014xna}; for reference, also the value $r=0.01$ is displayed.   }
\label{fig-1}
\vskip .2 cm
\end{figure}

We first start with the analysis based on the $\overline{\rm MS}$ top mass.
In fig.\ref{fig-1}, the (red and orange) lines spanning the $2\sigma$ range of $\alpha_3(m_Z)$, show 
the points in the ${\overline{ m}}_t(m_t)-m_H$ plane where the shallow SM false minima exists.
The thickness of the lines is associated to the theoretical uncertainty of the NNLO procedure, due to the matching of $\lambda$.
The latter are extremely close to the line marking the transition from stability to metastability (two degenerate vacua), so that it is not possible to distinguish them by eye, the difference being a few MeV for the top mass \cite{Masina:2012tz}.
The (green)  shaded horizontal region corresponds to the $1 \sigma$ range ${\overline{ m}}_t(m_t)=163.3 \pm 2.7 $ GeV, 
as calculated in ref. \cite{Alekhin:2012py}. 
The (pink) shaded vertical regions are the 1 and 2 $\sigma$ ranges of $m_H$ according to the PDG \cite{PDG}. 
The shallow false minimum configuration (and, more generally, any configuration close to it, like e.g. two degenerate minima) 
is thus compatible with the Higgs and top quark mass values, provided the latter is quite light.

For each shallow false minimum configuration the highness of the Higgs potential at the false minimum, $V(\chi_0)$, can be calculated and,
thanks to eq. (\ref{eq-M}), the same is true for $H_I$.
Let now consider the tensor to scalar ratio of cosmological perturbations. 
The amplitude of  density fluctuations in the observed Universe as seen by the CMB and Large-Scale structure data  
is parametrized by the power spectrum in $k$-space
\be
P_s(k)=\Delta_R^2 \left( \frac{k}{k_0} \right)^{n_S-1} \,\,,
\ee
where $\Delta_R^2$ is the amplitude at some pivot point $k_0$.
We consider the best-fit value from \cite{Ade:2013zuv}, 
$\Delta_R^2= (2.20 \pm 0.05) \times  10^{-9}$ 
 at $k_0=0.002 \,{\rm Mpc}^{-1}$.

In any inflationary model that can be analyzed through the slow-roll approximation, there is a  relationship between the scale of inflation, 
the amplitude of density perturbations, and the amount of gravity waves that can be produced:
\be 
\Delta_R^2 =  \frac{2}{3 \pi^2}  \frac{1}{r} \frac{ V(\chi_0) }{ M^4  }  \,\,.
\label{eq-r}
\ee
If inflation actually started from a SM shallow false minimum, then each point in the ${\overline{ m}}_t(m_t)-m_H$ plane has to
be associated with a specific value of $r$. 
This is done in fig.\ref{fig-1} , where the (blue) shaded bands represents the $1$ and $2$ 
$\sigma$ ranges of BICEP2 measurement $r=0.16\ ^{+0.06}_{-0.05}$ \cite{Ade:2014xna}; for reference, also the value $r=0.01$ is displayed.
Taking into account the theoretical error in the determination of the position of the false minimum (the thickness of the red and orange lines), 
the position of the (blue) $r$ bands is also uncertain by about $\pm 0.1$ GeV along the vertical axis.

One can see that, given the results from BICEP2, the shallow false minimum is a viable framework for models of Higgs inflation provided that: 
i) the Higgs mass is close to its upper 1-2 $\sigma$ range, more precisely between $126$ and $126.7$ GeV, 
or ii) both  $\alpha_3(m_Z)$ and the top mass ${\overline{ m}}_t(m_t)$ are quite small, say respectively close to the lower 1-2 $\sigma$ range (between  $0.1162$ and $0.1179$) and between $160.5$ and $161.5$ GeV.
 Clearly, the smaller is $r$, the more the three previous parameters can go in the direction of their central values. 
 It will thus be important to further improve the $r$ measurement in the future.

\begin{figure}[t!]
 \includegraphics[width=7.5cm]{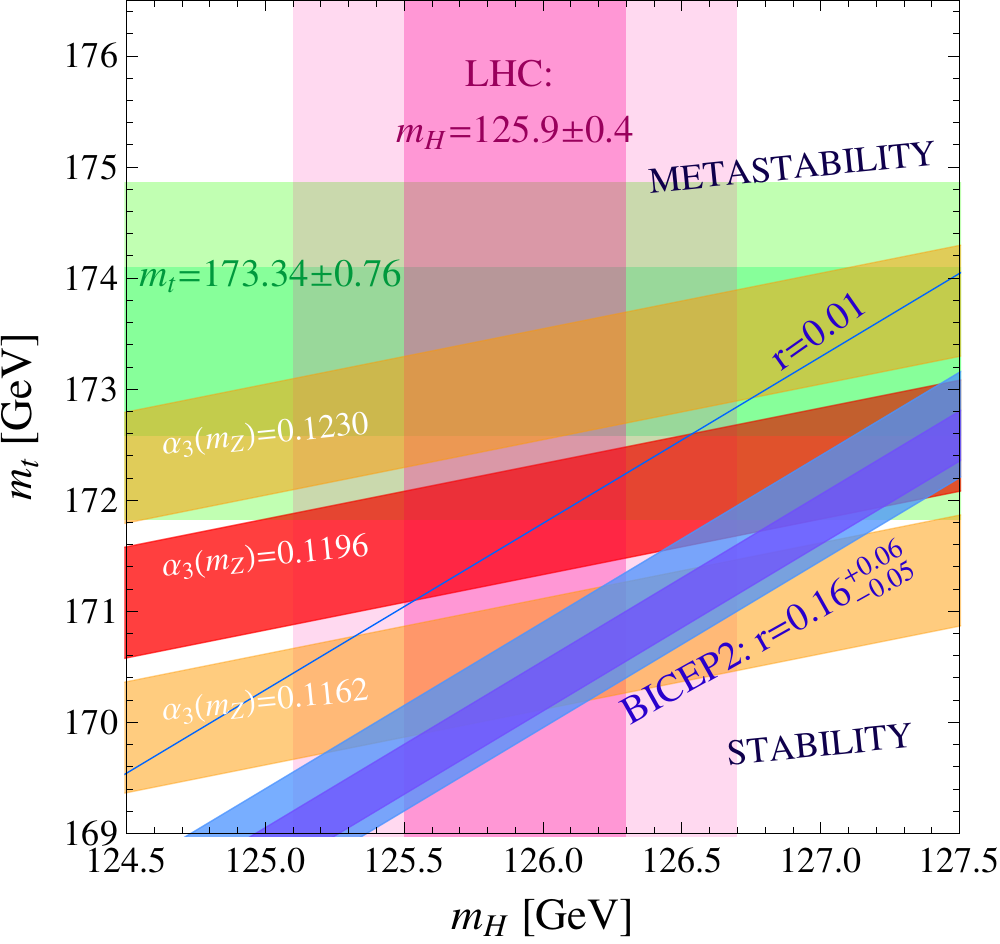}  
\caption{The (red and orange) lines corresponding to the quoted value of $\alpha_3(m_z)$ show the points where the SM false vacuum configuration is realized, their thickness being the theoretical error in the NNLO RGE procedure.
The (green)  shaded horizontal region corresponds to the $1$ and $1$ $\sigma$ ranges of the combined measurement 
by the ATLAS, CDF, CMS, D0 collaborations,
$m_t =173.34 \pm 0.76$ GeV \cite{ATLAS:2014wva}. 
The (pink) shaded vertical regions are the 1 and 2 $\sigma$ ranges of $m_H$ \cite{PDG}. 
The (blue) shaded bands are the $1$ and $2$ $\sigma$ ranges of BICEP2, $r=0.16\ ^{+0.06}_{-0.05}$ \cite{Ade:2014xna}; for reference, also the value $r=0.01$ is displayed.  }
\label{fig-2}
\vskip .2 cm
\end{figure}

The same conclusions can be drawn by making the analysis using the top pole mass, see fig. \ref{fig-2}.
As the theoretical error has to include also the uncertainty due to the matching of the top yukawa coupling, the thickness of the lines is bigger than with the previous method, and turns out to be about $\pm 0.5$ GeV in the vertical axis (see e.g. \cite{Degrassi:2012ry}). 
For the sake of clearness, we thus display only three lines, corresponding the the central and $2\sigma$ values of $\alpha_3(m_Z)$.
Taking into account the theoretical error in the determination of the position of the false minimum (the thickness of the red and orange lines), 
the position of the (blue) bands representing the BICEP2 result is also uncertain by a shift along the vertical axis of about $\pm 0.5$ GeV.
The (green) shaded horizontal regions show the 1 and 2 $\sigma$ ranges of the combined measurement by the ATLAS, CDF, CMS, D0 collaborations,
$m_t =173.34 \pm 0.76$ GeV \cite{ATLAS:2014wva}.
The (pink) shaded vertical regions are the 1 and 2 $\sigma$ ranges of $m_H$ according to the PDG \cite{PDG}. 

Also from fig. \ref{fig-2} one can see again that, for the shallow false minimum configuration, the BICEP2 results are compatible with the present ranges for $\alpha_3(m_Z)$, $m_H$, $m_t$.
In particular the overlapping requires: $m_t$ close to its lower  $2\sigma$ value,  $m_H$ close to its upper $2\sigma$ value,
$\alpha_3(m_Z)$ close to its central one.

Notice that the BICEP2 result at $1\sigma$, $r=0.16\ ^{+0.06}_{-0.05}$ \cite{Ade:2014xna},
corresponds to the range $V(\chi_0)=(1.8-2.2) \times 10^{16}$ GeV, and to $ H_I= (0.8-1.2) \times 10^{14}$ GeV.


Summarizing, we have argued that the present status of the measurements of $r$, $m_t$, $m_H$, $\alpha_3(m_Z)$,
is consistent with of the hypothesis \cite{Masina:2011aa} that inflation occurred in a SM shallow false vacuum at about $2\times 10^{16}$ GeV.
To account for the BICEP2 result at $1\sigma$, $r=0.16\ ^{+0.06}_{-0.05}$ \cite{Ade:2014xna}, 
in particular, the top quark mass should be close to its lower $2\sigma$ range, while the Higgs mass should be close to its $2 \sigma$ upper one,
see figs. \ref{fig-1} and \ref{fig-2}. 

It is intriguing that for such values of the top and Higgs masses, the quadratic divergences of the Higgs mass (corresponding to the running Veltman condition) cancel at the Planck scale, see fig. 2 of ref. \cite{Masina:2013wja}.

Future precision measurements of the top quark and Higgs masses will thus be crucial to further test the SM shallow false minimum inflationary scenario.

\section*{Acknowledgements}
We thank A. Notari and A. Strumia for useful discussions.


\end{document}